\documentclass[epj]{svjour}
\usepackage{graphics,epsfig,amsmath}
\begin{document}

\title{Inelastic neutron scattering study and Hubbard model description of the antiferromagnetic tetrahedral molecule Ni$_4$Mo$_{12}$}

\titlerunning{INS study and Hubbard model description of Ni$_4$Mo$_{12}$}

\author{J. Nehrkorn\inst{1} \and M. H\"{o}ck\inst{2} \and M. Br\"{u}ger\inst{3} \and H. Mutka\inst{4} \and J. Schnack\inst{2} \and O. Waldmann\inst{1}
}                     
%
\authorrunning{ J. Nehrkorn, M. H\"{o}ck, \emph{et al.}}
\institute{Physikalisches Institut, Universit\"{a}t Freiburg,
Hermann-Herder-Str. 3, D-79104 Freiburg, Germany \and
Universit\"{a}t Bielefeld, Fakult\"{a}t f\"{u}r Physik,
Postfach 100131, D-33501 Bielefeld, Germany \and
Universit\"{a}t Osnabr\"{u}ck, Fachbereich Physik, D-49069
Osnabr\"{u}ck, Germany \and Institut Laue-Langevin, 6 Rue
Jules Horowitz, BP 156, F-38042 Grenoble Cedex 9, France}

\date{Received: date / Revised version: date}

\abstract{ The tetrameric Ni(II) spin cluster Ni$_4$Mo$_{12}$ has been studied by INS. The data were analyzed
extensively in terms of a very general spin Hamiltonian, which includes antiferromagnetic Heisenberg interactions,
biquadratic 2-spin and 3-spin interactions, a single-ion magnetic anisotropy, and Dzyaloshinsky-Moriya interactions.
Some of the experimentally observed features in the INS spectra could be reproduced, however, one feature at 1.65 meV
resisted all efforts. This supports the conclusion that the spin Hamiltonian approach is not adequate to describe the
magnetism in Ni$_4$Mo$_{12}$. The isotropic terms in the spin Hamiltonian can be obtained in a strong-coupling
expansion of the Hubbard model at half-filling. Therefore detailed theoretical studies of the Hubbard model were
undertaken, using analytical as well as numerical techniques. We carefully analyzed its abilities and restrictions in
applications to molecular spin clusters. As a main result it was found that the Hubbard model is also unable to
appropriately explain the magnetism in Ni$_4$Mo$_{12}$. Extensions of the model are also discussed.
} 

\maketitle

\section{Introduction}\label{sec:introduction}

In the recent years molecular nanomagnets have attracted huge
interest because of their sometimes spectacular magnetic
properties. For instance, in molecules such as Mn$_{12}$ or
Fe$_8$ slow magnetic relaxation or even quantum tunneling of the
magnetization have been observed \cite{Mn12a,Mn12b,Mn12c}. A
general definition which includes all relevant possibilities is
difficult, but in most cases molecular nanomagnets consist of
magnetic metal ions with 3d shells and organic ligands. Also, in
most cases the magnetism is very well described by assuming
localized magnetic moments, such that metal ions with quenched
orbital angular momentum are described simply by their total
atomic spin $\hat{\vec{S}}_i$ ($i$ numbers the metal centers in
the molecule). This yields a spin Hamiltonian, which typically
includes a Heisenberg exchange term describing the magnetic
interactions between the metal centers in a molecule, a
zero-field-splitting (ZFS) term describing the magnetic
anisotropy due to the ligand-field effect, and a Zeeman term
accounting for the effects of an applied magnetic field
\cite{BCC:IC99}. However, also more complicated terms may be
relevant, such as Dzyaloshinsky-Moriya (DM) interactions, 3-spin
and 4-spin exchange interactions, or higher-order ligand-field
terms.

Today, molecular nanomagnets comprise a large number of magnetic
molecules, and it seems fair to say that the larger a molecule
is the more interesting it tends to be. Here larger means a
larger number $N$ of metal centers and/or larger spin lengths
$S_i$, i.e., a larger dimension of the Hilbert space. A larger
dimension typically gives rise to a richer structure of the
energy spectra and wave functions, and wherewith to more
interesting magnetic phenomena. Also, technical aspects such as
the calculation of magnetic observables from the spin
Hamiltonian become quickly difficult, which establishes an
interesting challenge by itself.

In this context, i.e., understanding the magnetic excitations in large spin clusters, interesting molecules are for
instance the antiferromagnetic (AFM) molecular wheels, in which 6 to 18 exchange-coupled metal ions form rings
\cite{wheels1,wheels2,wheels2b,wheels3,wheels4,wheels5,wheels6,wheels6b,wheels7,wheels8,wheels8b,wheels9,wheels10,wheels11}.
These wheels enabled the observation of phenomena such as the rotation of the N\'eel vector \cite{nrot1}, which is
related to the tower of states in antiferromagnetics \cite{nrot2,nrot3}, quantized AFM spin waves \cite{nrot1,waves1},
or in the larger wheels quantum tunneling of the N\'eel vector and the associated quantum interference effects
\cite{nvt1,nvt2,nvt3,nvt4}. Numerically, for hexanuclear wheels the magnetism is easily calculated exactly from the
spin Hamiltonian \cite{num1,num2}, and in part also for CsFe$_8$ with a dimension of the Hilbert space of 1.679.616
\cite{wheels7}. However, for the Fe$_{18}$ wheel \cite{wheels10}, with a Hilbert space as large as $\approx 10^{14}$,
the interpretation of the experiments requires advanced approximate techniques, which build on physical insight
\cite{nvt4}. Another interesting molecule is the Keplerate Mo$_{72}$Fe$_{30}$ \cite{fe30}, in which 30 Fe(III) ions
occupy the symmetric sites of an icosidodecahedron giving rise to strong magnetic frustration effects \cite{KMS:CCR09}.
Phenomena such as plateaus in the magnetization at 1/3 magnetization due to competing spin phases
\cite{spinphase,RLM:PRB08} or the presence of low-lying singlets \cite{SSR:JMMM05} were observed or predicted. The
Hilbert space is huge, of dimension $\approx 10^{23}$, hence a detailed understanding of the magnetism in this cluster
is obviously difficult \cite{fe30bands,ExS:PRB03}.

In contrast to this trend to larger molecules, sometimes even very small magnetic molecules, which at first sight would
be discarded as trivial because of their small Hilbert space (which implies that ``everything'' can be calculated
easily), may exhibit striking magnetic behavior. For instance, the Fe(III) dimer molecule [Fe$_2$F$_9$(Et$_4$N)$_3$]
exhibits unusual magnetization dynamics at low temperatures with signatures of quantum tunneling of the magnetization
\cite{fe2a,fe2b,fe2c}, the unusual magnetic behavior in the Cu(II) tetrahedron [Cu$_4$OCl$_6$daca$_4$] was associated
to phonon interactions \cite{Cu4} and the Ni(II) single-molecule magnet [Ni$_4$((hmp)(t-BuEtOH)Cl)$_4$] allowed one to
realize quantum superpositions of high-spin states \cite{ni4smm1,ni4smm2,ni4smm3}.
\begin{figure}
\begin{center}
\includegraphics{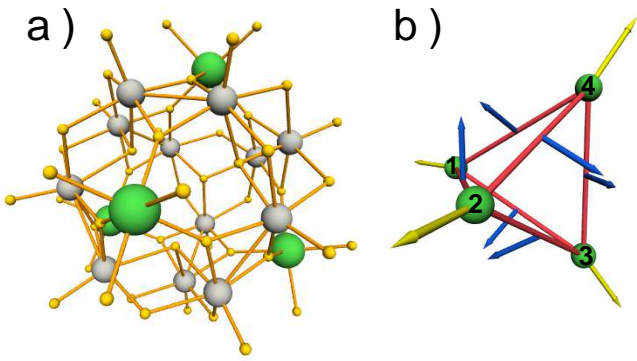}
\end{center}
\caption{
  (a) Ball-and-stick representation of the magnetic core of Ni$_4$Mo$_{12}$. Big green balls: Ni(II) ions, grey balls: Mo, small golden balls: O.
  H atoms were omitted for clarity. (b) Assumed coupling paths (red/dark sticks), orientations of the local anisotropy axes (yellow/light arrows)
  and DM vectors (blue/dark arrows) of Ni$_4$Mo$_{12}$.
}
\label{fig:magcore}
\end{figure}

In this work we will study the tetrameric Ni(II) spin cluster
[Mo$_{12}$O$_{30}$($\mathrm{\mu}_2$-OH)$_{10}$H$_2$(Ni(H$_2$O)$_3$)$_4$], or Ni$_4$Mo$_{12}$ in short, which is another
such example \cite{Ni4synthese}. In Ni$_4$Mo$_{12}$ four Ni(II) ions occupy the vertices of an almost perfect
tetrahedron and exhibit AFM nearest-neighbor Heisenberg interactions. The core of Ni$_4$Mo$_{12}$ is shown in
Fig.~\ref{fig:magcore}(a). The magnetism of this ``simple'' cluster should be unspectacular and easy to describe. Due
to the AFM Heisenberg interactions, the ground state should belong to total spin $S=0$, followed by a sequence of
$S=1,2,3,4$ states with energies obeying the Land\'e rule $E(S) = \frac{1}{2}\Delta S(S+1)$. In an applied magnetic
field, this should give rise to a series of level crossings (LCs), where the ground state changes from $S=0$ to $S=1$,
$S=1$ to $S=2$, and so on. This sequence of LCs should in turn be detected, at low temperatures, in the magnetization
curve as a sequence of steps at regular fields $B_{n} = n \Delta/(g \mu_B)$, with $n=1,2,3,...$ \cite{wheels1}. For
Ni$_4$Mo$_{12}$ magnetization steps have indeed been observed, but at fields of 4.5, 8.9, 20.1, and 32~T, which is
incompatible with the Heisenberg picture \cite{Ni4Schnack}. This discrepancy could not be resolved by additionally
introducing ZFS and biquadratic exchange terms in the spin Hamiltonian, and a magnetic-field dependence of the exchange
and ZFS parameters was hence proposed \cite{Ni4Schnack}. Subsequently, Kostyuchenko pointed out that 3-spin
interactions should not be neglected \cite{Kost}. Such terms can either originate from spin-phonon interactions as
suggested in Ref. \cite{Ni4Schnack} or from electron delocalization as described by a Hubbard model \cite{Kost}.
Indeed, the inclusion of such terms allowed Kostyuchenko to reproduce the field positions of the magnetization steps.
Also, starting from a Hubbard model at half-filling, the strengths of the Heisenberg, biquadratic, and 3-spin
interaction terms in the spin Hamiltonian were obtained. According to this result, the biquadratic and 3-spin
interactions should be related by a factor of 2 (in our units), which was found to be in agreement with experiment as
determined from a fit of the model to the experimental field positions. This result was interpreted as to indicate the
superiority of the Hubbard model for Ni$_4$Mo$_{12}$. More recently, Klemm and Efremov presented an extensive analysis
of the magnetization in tetrameric spin clusters based on a general spin Hamiltonian \cite{Klemm}. However,
unfortunately, the symmetry case relevant for Ni$_4$Mo$_{12}$ was not considered.

In order to better understand the unusual magnetism in Ni$_4$Mo$_{12}$, we undertook inelastic neutron scattering (INS)
experiments as well as a more detailed analysis of the Hubbard model, which we will present in this work. INS is
renowned for its ability to study exchange splittings in magnetic clusters directly
\cite{nrot1,superINS1,superINS2,superINS3,SFA:PRB07,WKD:JSSC03,MDM:CAEJ06,HRZ:PRL05}. Our extensive analysis of the
data in terms of a very general spin Hamiltonian will provide insight into the importance of the various interaction
terms, but we could not find a parameter set which reproduces all key aspects of the data. This is certainly an
unsatisfying outcome, but emphasizes the unconventionality of the magnetism in Ni$_4$Mo$_{12}$. It also suggests to
study models going beyond the spin Hamiltonian approach, such as the Hubbard model \cite{Hubbard}, which allows for
mobile electrons, i.e. itinerant magnetic moments. At half-filling the electron mobility is governed by the Hubbard-$U$
parameter, and the model of localized moments is recovered in the limit of large $U$ (strong-coupling limit). Our work
on the Hubbard model is the most careful application of it to a molecular spin cluster to date. However, in contrast to
Ref.~\cite{Kost}, we find using various techniques that the Hubbard model is not adequate for describing the magnetism
in Ni$_4$Mo$_{12}$. We will clarify this discrepancy, and as a byproduct resolve some errors in previous works.

We like to mention that a Hubbard model description of a
magnetic molecule could be of interest by its own. For example,
the low-energy spectrum might contain additional levels not
present in a pure spin model, which may be crucial for an
appropriate theoretical description of the magnetism in a
molecule. As mentioned earlier, also biquadratic multiple-spin
interactions such as 2-, 3-, and 4-spin terms appear in the
strong-coupling limit of the Hubbard model \cite{tU}. There is
thus the prospect of identifying electron delocalization as
suggested by Density Functional Theory calculations
\cite{FNL:JACS96} and modeled by the Hubbard Hamiltonian as the
physical mechanism leading to such interactions and giving a
better understanding of them.

The remainder of this article is organized as follows. In
section~\ref{sec:experiment} the experiments and a first
qualitative analysis are described. Section~\ref{sec:ham}
presents the phenomenological spin Hamiltonian that is used in
section~\ref{sec:analysis} to analyze the data with respect to
various parametrizations. Section~\ref{sec:hubbard} discusses a
possible description of Ni$_4$Mo$_{12}$ in terms of a Hubbard
model. The article closes with conclusions.

\section{Experiments and Qualitative Analysis}\label{sec:experiment}

A fully-deuterated powder sample of Ni$_4$Mo$_{12}$ was used for
the INS experiment. For details on synthesis and molecular
structure see \cite{Ni4synthese}. The INS data were measured on
the direct time-of-flight spectrometer IN5 at the Institut
Laue-Langevin (ILL, Grenoble, France). Spectra were recorded at
temperatures of 2.4, 9.3, and 23 K for an incident neutron
wavelength of $\lambda$ = 5.0~{\AA}. The energy resolution at
the elastic peak was 118~$\mathrm{\mu}$eV. The data were
corrected for detector efficiency via a vanadium standard, and
spectra were summed over all detector banks.

\begin{figure}
\begin{center}
\includegraphics{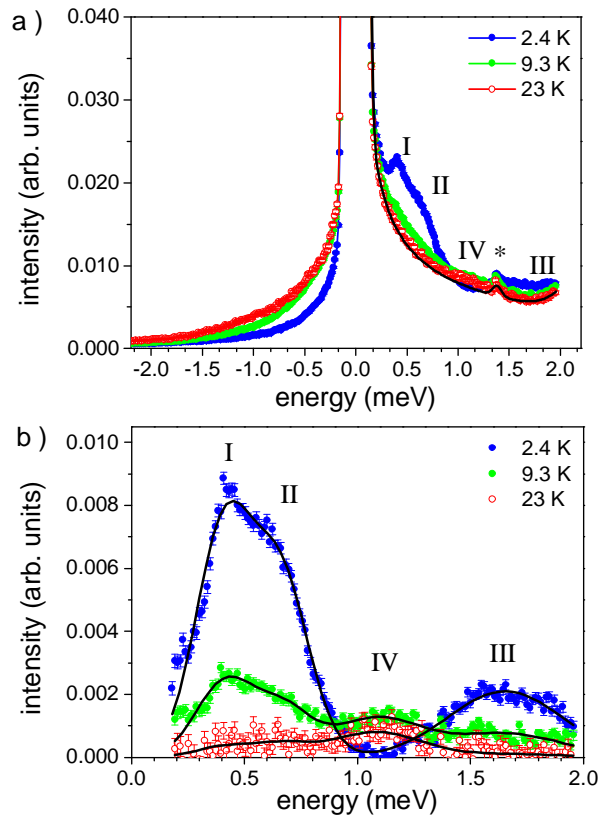}
\end{center}
\caption{
(a) INS spectra of Ni$_4$Mo$_{12}$ at the indicated temperatures. The black line represents the background we chose (see text).
(b) The INS spectra with the background subtracted. The black lines are multi-Gauss fits with parameters given in Table~\ref{tab1}.
}
\label{fig:ins_roh}
\end{figure}

Fig.~\ref{fig:ins_roh}(a) presents the INS spectra. Four features can be identified on the neutron energy-loss side. At
2.4~K two strong peaks at about 0.4 and 0.65~meV (peaks I and II henceforth) are clearly visible. Their intensity
decreases significantly with increasing temperature. Hence, these peaks correspond to cold transitions and are magnetic
in origin. At a higher energy of about 1.7~meV (peak III) additional intensity is also visible at the lowest
temperature. This broad feature is hence also assigned to a cold magnetic transition. At intermediate energies of about
1.1~meV (peak IV) additional intensity appears at higher temperatures, which we thus assign to a hot magnetic peak. The
sharp feature at 1.4~meV marked by an asterisk in Fig.~\ref{fig:ins_roh}(a) is a spurion. The scattering intensity on
the neutron energy-gain side increases with increasing temperature, in accord with general expectations and in
agreement with the above assignment of the features I to IV, but is pretty featureless, which is consistent with the
reduced resolution for up scattering in direct TOF instruments. We hence focus on the data on the energy-loss side in
the following.

In order to be able to better analyze the features, we
subtracted a background curve from the neutron energy-loss data
[shown as a black solid line in Fig.~\ref{fig:ins_roh}(a)],
which we generated from an educated guess. Overall the true
contribution from the nonmagnetic scattering should be
approximated well by our curve, but it might of course be
incorrect regarding finer details. The background corrected
spectra are plotted in Fig.~\ref{fig:ins_roh}(b), showing the
four features I to IV more clearly. Multi-Gauss fits to the
peaks were done for each temperature, and the obtained peak
positions, widths, and intensities are compiled in
Table~\ref{tab1}.

\begin{table}
\caption{Results of a multi-Gauss fit analysis of the INS spectra of Ni$_4$Mo$_{12}$ discussed in the text. The peak
positions (in meV), line widths (FWHM, in meV), and peak intensities (in $10^{-3}$ arb. units) of the fitted Gauss
curves are listed.} \label{tab1}
\begin{tabular}{cccccc}
\hline
& I & II & III & IV  \\
\hline
energy        & 0.408(3) & 0.663(4) & 1.65(1) & 1.08(5)\\
FWHM          & 0.15     & 0.15     & 0.30(2) & 0.23(2)\\
intensity @ 2.4~K & 2.26(4)  & 1.74(4)  & 1.33(5) & 0\\
intensity @ 9.3~K & 0.73(2)  & 0.47(2)  & 0.48(3) & 0.60(3)\\
intensity @ 23~K  & 0.11(2)  & 0.12(2)  & 0.07(3) & 0.39(2)\\
\hline
\end{tabular}
\end{table}

\begin{figure}
\begin{center}
\includegraphics{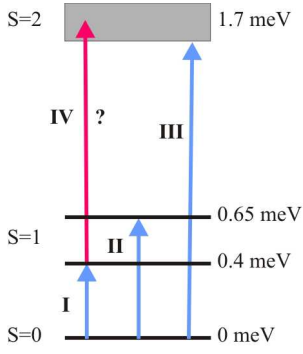}
\end{center}
\caption{
Energy level structure as deduced from the INS data. The assignment of the levels to total spin quantum numbers is discussed in the text.
}
\label{fig:ins_spectrum}
\end{figure}

The INS data indicates the energy level structure shown in Fig.~\ref{fig:ins_spectrum}. The cold peaks I, II, and III
are assigned to transitions from the ground state to three levels at 0.4, 0.65, and 1.7~meV. The energy of peak IV
matches well the gap between the two lowest excited levels and the level at 1.7~meV, which suggests to assign it
accordingly. However, the temperature dependence of its intensity is maximal at around 10-15~K, such that this peak
should in fact originate from another higher lying level. The line widths of peak I and II are somewhat larger but
close to the experimental resolution, which suggests that they are made up of single transitions (broadened by e.g. $J$
strain). However, peak III is significantly broadened, which indicates that a band of energy levels exists at 1.7~meV.
The width of peak IV is also larger than the experimental resolution, but in view of the weakness of the signal
conclusions are not obvious.

A first attempt towards the interpretation of the INS data can
be made by comparison to the magnetization data. In the
magnetization curve at low temperatures four steps were observed
at fields of 4.5, 8.9, 20.1, and 32~T \cite{Ni4Schnack}. In a
Heisenberg-exchange picture these emerge from the Zeeman
splitting of the lowest multiplets for each total spin $S = 0,
1, 2, 3, 4$, as described in the introduction. Since the Zeeman
splitting of the involved states is known to be $- g \mu_B S B$,
the energy of the lowest spin states in zero field can directly
be determined from the magnetization steps. This yields that in
zero field the lowest $S = 1,..., 4$ levels should be at
energies of 0.6, 1.7, 4.4, and 8.5~meV (the ground state is
$S=0$). Comparing these values to the INS findings immediately
suggests to assign the peaks I and II to the $S=1$ level, with a
ZFS of its $M=0$ and $M=\pm1$ components of 0.25~meV, and peak
III to the $S=2$ level, with a weaker ZFS giving rise to an
enhanced line width. This assignment to spin levels is also
indicated in Fig.~\ref{fig:ins_spectrum}.

Energy wise, this interpretation is satisfying as it is
perfectly consistent with the observed magnetization curve.
However, it is inconsistent with basic rules governing INS
intensity in two instances: (1) Peak III would correspond to a
transition from a $S = 0$ to a $S = 2$ level, which is forbidden
by the INS selection rule $\Delta S = 0, \pm1$. (2) The $S=1$
levels would already be thermally populated at a temperature of
2.4~K (we estimate a thermal population of ca. 15\%) and
significant scattering intensity should be observed at this
temperature, in disagreement with experiment.  The above
interpretation has thus to be abandoned as too simple. The
behavior of peak III will in fact turn out to be a major
obstacle for all the models discussed in this work.

\section{Phenomenological Spin Hamiltonian}\label{sec:ham}

In view of the findings in previous works
\cite{Ni4Schnack,Kost,PhdBrueger} and the above INS data, the
spin Hamiltonian for Ni$_4$Mo$_{12}$ must include additional
non-Heisenberg interactions. Here we consider the interactions
\begin{subequations}\label{eq:1}
\begin{eqnarray}
  \widehat{\mathcal H}_{ex} &=& -J \sum_{i<j}^4 \hat{\vec{S}}_i \cdot \hat{\vec{S}}_j  \label{eq:heisenberg},\\
  \widehat{\mathcal H}_{2} &=& -J_2\sum_{i<j}^4 (\hat{\vec{S}}_i \cdot \hat{\vec{S}}_j)^2 \label{eq:2spin},\\
  \widehat{\mathcal H}_{3} &=& -J_3\sum_{i\neq j \neq k}^4 (\hat{\vec{S}}_i \cdot \hat{\vec{S}}_j)(\hat{\vec{S}}_j \cdot \hat{\vec{S}}_k)\label{eq:3spin}, \\
  \widehat{\mathcal H}_{DM} &=& -\sum_{i<j}^4  \vec{d}_{ij} \cdot (\hat{\vec{S}}_i \times \hat{\vec{S}}_j )\label{eq:DM},\\
  \widehat{\mathcal H}_{ZFS} &=& \sum_{i=1}^4 \hat{\vec{S}}_i \cdot \pmb{D}_i \cdot \hat{\vec{S}}_i\label{eq:ZFS},
\end{eqnarray}
\end{subequations}
where $\widehat{\mathcal H}_{ex}$ describes the AFM
nearest-neighbor Heisenberg interactions ($J<0$),
$\widehat{\mathcal H}_{2}$ and $\widehat{\mathcal H}_{3}$ are
the isotropic fourth-order exchange terms linking 2 and 3 spins,
respectively, $\widehat{\mathcal H}_{DM}$ describes the
antisymmetric DM interactions \cite{DM1,DM2}, and
$\widehat{\mathcal H}_{ZFS}$ describes the ZFS due to the local
on-site magnetic anisotropy. To avoid confusion later on, we
repeat that $\hat{\vec{S}}_i$ denotes the spin operator of the
$i$th Ni(II) ion ($S_i=1$ for all $i$).

In principle, if going to fourth order in the isotropic exchange then also a term linking 4 spins should be included.
Its general structure is $(\hat{\vec{S}}_i \cdot \hat{\vec{S}}_j)(\hat{\vec{S}}_k \cdot \hat{\vec{S}}_l)$ with four
different indices $i, j, k$, and $l$. According to Ref. \cite{Kost}, this 4-spin contribution vanishes in the case of
Ni$_4$Mo$_{12}$; we hence disregarded it in our analysis. Our discussion in section~5 will show that there is no reason
to assume that the 4-spin interactions cancel out, but we checked that neglecting them does not affect our general
conclusions.

Because of the nominal high symmetry of Ni$_4$Mo$_{12}$ and to
avoid a further increase in the number of parameters, we assumed
that the isotropic coupling constants $J$, $J_2$, and $J_3$ are
equal for each coupling path. The Hamiltonian $\widehat{\mathcal
H}_{ex} + \widehat{\mathcal H}_{2} + \widehat{\mathcal H}_{3}$
can be solved analytically by using the vector-coupling rules of
angular momenta (Kambe's method) \cite{Kost,Klemm}. However, the
eigenvalues provided in Refs.~\cite{Kost} and \cite{Klemm}
disagree. We hence performed systematic test calculations using
exact numerical diagonalization, which confirmed the eigenvalues
in Ref.~\cite{Kost}.

According to the DM rules \cite{DM2}, the DM vectors
$\vec{d}_{ij}$ should vanish for a perfectly tetrahedral
cluster.  However, a slight distortion from this high symmetry
allows for non-zero DM interactions, which can in fact become
significant \cite{DM2,Ni4Kirchner}. Assuming a $S_4$ symmetry,
the orientations of the $\vec{d}_{ij}$ shown in
Fig.~\ref{fig:magcore}(b) result, with equal lengths of the DM
vectors for all coupling paths. We therefore write $\vec{d}_{ij}
= d \vec{e}_{ij}$ with appropriate unit vectors $\vec{e}_{ij}$.

The on-site anisotropy terms may also be effective in a perfectly tetrahedral cluster, but the local anisotropy tensors
are obviously severely constrained in their orientation and magnitude by symmetry. To avoid over-parametrization, we
assumed an axial local anisotropy on each Ni site, each characterized by a tensor $\pmb{D} = D \texttt{diag}(
-\frac{1}{3}, -\frac{1}{3}, \frac{2}{3})$ in the local coordinate frame. The local coordinate frames are related to the
cluster coordinate frame by rotation matrices $\pmb{R}(\alpha_i, \vartheta_i, \varphi_i)$, with the Euler angles
$\alpha_i$, $\vartheta_i$, and $\varphi_i$ \cite{eulerbook}. In the cluster frame the local anisotropy tensors hence
become $\pmb{D}_i = \pmb{R}(\alpha_i, \vartheta_i, \varphi_i) \cdot \pmb{D} \cdot \pmb{R}^T(\alpha_i, \vartheta_i,
\varphi_i)$. The local $z$ axes were chosen to point radially outwards, as shown in Fig.~\ref{fig:magcore}(b). In
principle, any configuration of the axes which is related to this one by a global rotation would also satisfy the
tetrahedral symmetry. However, such a global rotation does not affect the magnetic properties of powder samples (we
neglect the weak interplay of local anisotropy and DM interactions). We mention that we also considered an additional
orthorhombic on-site anisotropy, i.e., a local ZFS tensor of the form $\pmb{D} = \texttt{diag}( -\frac{1}{3}D + E,
-\frac{1}{3}D-E, \frac{2}{3}D)$, but this did not lead to a significant improvement or further insight. We hence
disregarded this term.

In an applied magnetic field, the Zeeman term
\begin{equation}
\widehat{\mathcal H}_{B} = \mu_B g \vec{B} \cdot \sum_{i=1}^{4} \hat{\vec{S}}_i  \label{eq:zeeman}
\end{equation}
is additionally present, where $g$ is typically on the order of
2.3 for Ni(II) ions. We have neglected an anisotropy of the $g$
tensor, as it is expected to be small and irrelevant.

For a given spin Hamiltonian, the magnetization curve as well as
the INS spectra were calculated numerically from the eigenpairs
obtained from a full exact numerical diagonalization. For the
magnetization curves, the powder average was done by numerically
averaging over a grid of magnetic field orientations. The powder
INS spectra were calculated using the formulae given in
Refs.~\cite{INS1,INS2}.

\section{Analysis}\label{sec:analysis}
\subsection{Heisenberg exchange plus on-site magnetic anisotropy}

\begin{figure}
\begin{center}
\includegraphics{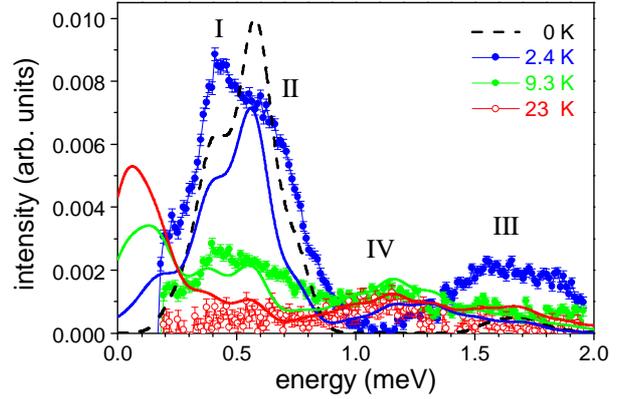}
\end{center}
\caption{
Experimental (dots) and simulated (lines) INS spectra for
$\widehat{\mathcal H} = \widehat{\mathcal H}_{ex} + \widehat{\mathcal H}_{ZFS}$ with $J = -6.5$~K and $D =3.25$~K.
}\label{fig:smix}
\end{figure}

The qualitative interpretation of the INS spectra in terms of a dominant Heisenberg interaction has shown that peak III
would violate the INS selection rule $\Delta S = 0, \pm1$. However, for Ni(II) ions it is well known that they may
exhibit on-site anisotropies as large as several ten K \cite{Boca}. Hence, the possibility arises that in
Ni$_4$Mo$_{12}$ the anisotropy constant $D$ is on the order of or even larger than the exchange coupling $J$, which
would give rise to strong mixing of spin levels, such that $S$ would cease to be a good quantum number
\cite{INS2,Liviotti,Datta}. Then also $\Delta S = 0, \pm1$ would cease to be a good selection rule. In a hope that this
mechanism may explain the significant INS intensity of peak III, we did extensive simulations for the Hamiltonian
$\widehat{\mathcal H} = \widehat{\mathcal H}_{ex} + \widehat{\mathcal H}_{ZFS}$, scanning the whole parameter regime
for $J$ and $D$. This model accounts for the two most significant interaction terms in Ni$_4$Mo$_{12}$. In order to be
reasonably consistent with the experimental data (magnetization and INS), we observed that $D$ cannot be very large,
such that the mixing effect is not very strong.  Accordingly, the simulated intensity for peak III is way too weak. The
best simulation of the INS spectra was obtained for $J = -6.5$~K and $D = 3.25$~K, see Fig.~\ref{fig:smix}. The peak
positions are well reproduced, but the discrepancies as regards the scattering intensity are obvious. Most noteworthy,
the intensity of peak III is strongly underestimated and exhibits a totally wrong temperature dependence. We couldn't
find any parameter set for which these discrepancies did not arise.

\subsection{The models of Schnack et al.}

\begin{figure}
\begin{center}
\includegraphics{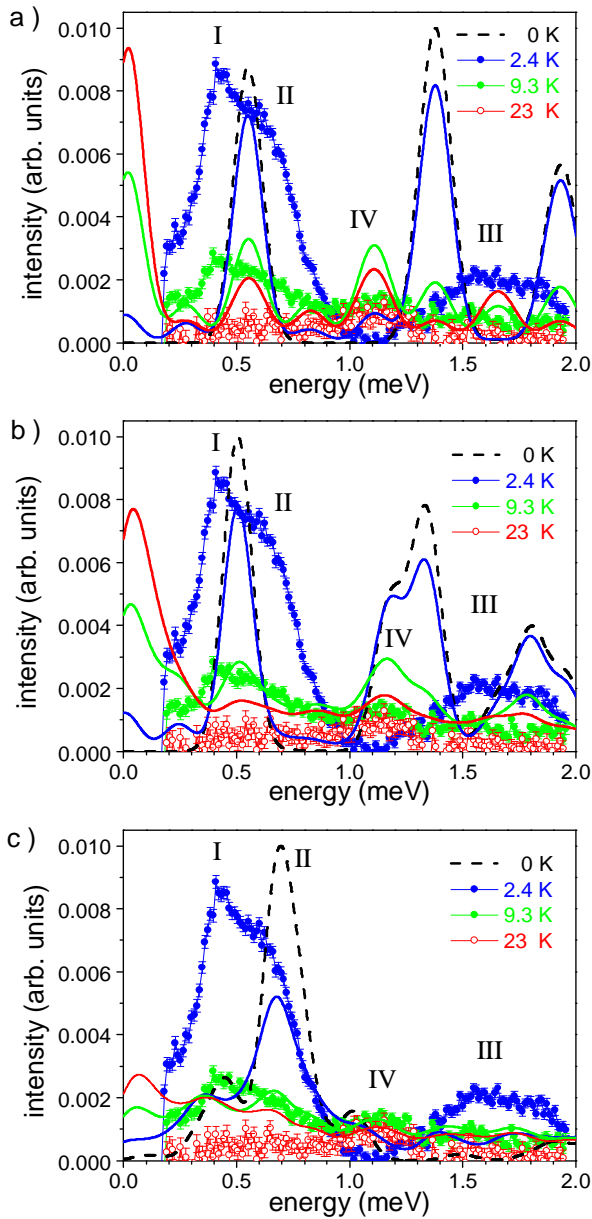}
\end{center}
\caption{
Experimental (dots) and simulated (lines) INS spectra for the three models
suggested in Ref. \cite{Ni4Schnack}, see also text and Table~\ref{tab2}.}
\label{fig:schnack}
\end{figure}

\noindent \begin{table}
\caption{Parameters of the three models discussed in Ref. \cite{Ni4Schnack}.}
\label{tab2}
\begin{tabular}{cccccc}
\hline
model & $J^a$ & $J^b$ & $J^a_2$ & $J^b_2$ & $D$  \\
\hline
a       & -6.4~K & -6.4~K & 3.2~K & 0~K & -1.0~K\\
b       & -6.4~K & -6.2~K & 3.0~K & 0~K & -3.2~K\\
c       & -8.4~K & -6.2~K & 0.32~K & 0.78~K & -8.9~K\\
\hline
\end{tabular}
\end{table}

Based on the magnetic susceptibility and magnetization curves, three models were suggested for Ni$_4$Mo$_{12}$ in Ref.
\cite{Ni4Schnack}. In these models additionally a structural distortion was accounted for via different coupling
constants for the isotropic exchange, $J_{12}=J_{23}=J_{13} = J^a$ and $J_{14}=J_{24}= J_{34} = J^b$, and similarly for
the 2-spin interactions, introducing parameters $J_2^a$ and $J_2^b$. The parameters of the three models are reproduced
in Table~\ref{tab2} (in our units), where the suggested magnetic field dependence of some of the parameters is
irrelevant as our INS experiment was done in zero field. The simulated INS spectra are compared to the experimental
data in Fig.~\ref{fig:schnack}. Unfortunately, none of these models can reproduce the experimental INS spectra. The
effects of DM interactions and/or tilted anisotropy tensors should be even weaker (the effect of DM interactions will
be discussed in more detail in section 4.4). Hence, a deviation from the assumed tetrahedral symmetry is apparently not
at the heart of the unusual magnetism in Ni$_4$Mo$_{12}$.

\subsection{The model of Kostyuchenko}
\label{sec:kostyuchenko}

\begin{figure}
\begin{center}
\includegraphics{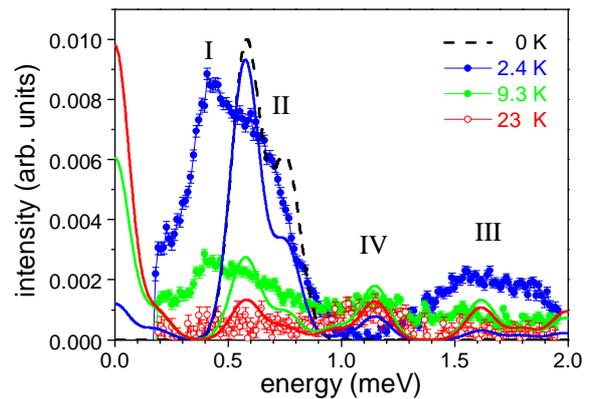}
\end{center}
\caption{
Experimental (dots) and simulated (lines) INS spectra for the model suggested in Ref. \cite{Kost},
$\widehat{\mathcal H} = \widehat{\mathcal H}_{ex} + \widehat{\mathcal H}_{2} + \widehat{\mathcal H}_{3}$ with $J = -8.82$~K, $J_2 = -1.08$~K, and $J_3 = -0.53$~K.}
 \label{fig:kost}
\end{figure}

Kostyuchenko showed recently that the positions of the steps in the magnetization curve can be reproduced very well by
the model $\widehat{\mathcal H} = \widehat{\mathcal H}_{ex} + \widehat{\mathcal H}_{2} + \widehat{\mathcal H}_{3} +
\widehat{\mathcal{H}}_B$, which in particular includes the 3-spin interactions \cite{Kost}. He also derived this spin
model from the strong-coupling limit of a Hubbard model at half-filling, and found that the coupling strengths $J$,
$J_2$, and $J_3$ are not independent. Unfortunately, as our analysis in section~5 will reveal, the obtained relations
are erroneous. Nonetheless, used as a phenomenological model, $\widehat{\mathcal H} = \widehat{\mathcal H}_{ex} +
\widehat{\mathcal H}_{2} + \widehat{\mathcal H}_{3} + \widehat{\mathcal{H}}_B$ very successfully describes the
magnetization data for the parameters $J = -8.82$~K, $J_2 = -1.08$~K, and $J_3 = -0.53$~K which were obtained from a
fit to the experimentally determined crossing fields \cite{Kost}. However, this model does not reproduce the
experimental INS spectra well, see Fig.~\ref{fig:kost}. The peak positions may be considered to be acceptable in view
of the simplicity of the model, but we again observe the same discrepancy as before, i.e., the intensity of peak III is
strongly underestimated and exhibits a wrong temperature dependence.

\begin{figure}
\begin{center}
\includegraphics{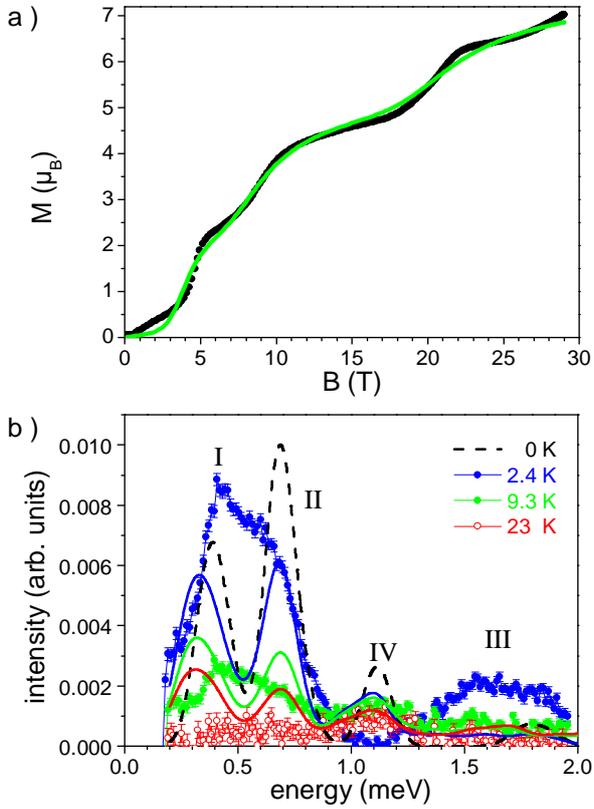}
\end{center}
\caption{
(a) Experimental (black dots) and simulated (green line) magnetization curve and (b) experimental (dots) and simulated (lines) INS spectra for
$\widehat{\mathcal H} = \widehat{\mathcal H}_{ex} + \widehat{\mathcal H}_{2} + \widehat{\mathcal H}_{3} + \widehat{\mathcal H}_{ZFS} + \widehat{\mathcal H}_{B}$
with $J = -9.45$~K, $J_2 = -1.41$~K, $J_3 = -0.66$~K, $D = 6.26$~K, and $g = 2.35$.}
\label{fig:mfit}
\end{figure}

Because ZFS can be significant for Ni(II) ions, we extended the model to $\widehat{\mathcal H} = \widehat{\mathcal
H}_{ex} + \widehat{\mathcal H}_{2} + \widehat{\mathcal H}_{3} + \widehat{\mathcal H}_{ZFS} + \widehat{\mathcal{H}}_B$
and ran a fit to the data, in which we used the magnetization and INS data simultaneously. The resulting best-fit
curves are shown in Fig.~\ref{fig:mfit}. The magnetization is well reproduced, but the fit to the INS data shows again
the ``peak III discrepancy''. Moreover, now also peak IV is incorrectly reproduced both as regards its intensity and
temperature dependence. Apparently, also this extended model cannot satisfactorily account for peak III.

\subsection{Systematic scan of the parameter regime}

In a last effort to explain the magnetism in Ni$_4$Mo$_{12}$ in terms of a spin Hamiltonian we systematically
considered the Hamiltonian $\widehat{\mathcal H} = \widehat{\mathcal H}_{ex} + \widehat{\mathcal H}_{2} +
\widehat{\mathcal H}_{3} + \widehat{\mathcal H}_{ZFS} + \widehat{\mathcal H}_{DM}$, which includes all terms considered
so far [see Eqs.~(\ref{eq:1})]. We started from simulations of the INS spectra for the Heisenberg Hamiltonian,
Eq.~(\ref{eq:heisenberg}), and then extended the model by including the other terms step by step. In every step, the
influence of the newly added term was analyzed. We will not discuss all details of our findings here, but just mention
some key observations.

\begin{figure*}
\begin{center}
\includegraphics{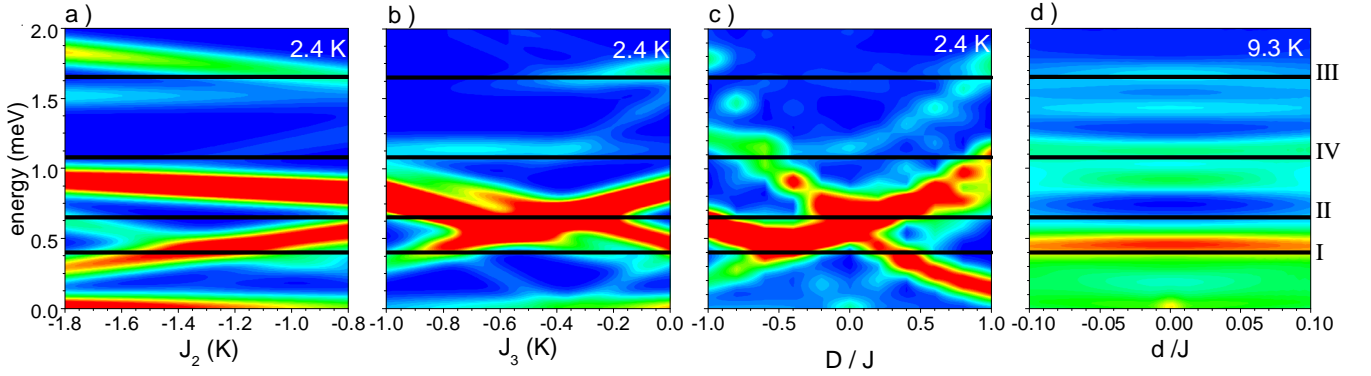}
\end{center}
\caption{
Simulated INS intensity as function of energy and a selected magnetic parameter. The strength of the scattering intensity is represented by the color (red = strong, blue = weak).
(a) Intensity vs. energy and $J_2$ at 2.4~K.
(b) Intensity vs. energy and $J_3$ at 2.4~K ($J_2 = -1.08 \, \text{K}$).
(c) Intensity vs. energy and $D$ at 2.4~K ($J_3 = -0.53 \, \text{K}$).
(d) Intensity vs. energy and $d$ at 9.3~K ($D = 3.5 \, \text{K}$).
As a guide to the eyes, the black lines indicate the energies of the measured peaks I, II, III, and IV.}
\label{fig:effects}
\end{figure*}

With the Heisenberg term, Eq.~(\ref{eq:heisenberg}), alone, no agreement between data and simulation could be obtained.
We hence chose $J = -8.82$~K. The effects of the further terms on the INS spectra are presented in
Fig.~\ref{fig:effects}. Each panel presents the INS intensity at one temperature, plotted as function of energy and the
magnetic parameter under consideration, with the scattering intensity represented by color. The dependence of the INS
intensity at 2.4~K on the strength of the 2-spin interaction $J_2$ is shown in Fig.~\ref{fig:effects}(a). Two
pronounced peaks at ca. 0.5 and 0.8~meV and a weak peak at ca. 1.8~meV are visible. They are close to the experimental
energies of features I, II, and III, but exhibit a rather weak dependence on $J_2$. Hence, the 2-spin interaction is
important, but its exact strength is not constrained much. We chose $J_2 = -1.08$~K. The simulations with additional
3-spin interactions are shown in Fig.~\ref{fig:effects}(b). The 3-spin interactions have a pretty strong effect on the
INS spectrum. The two low-energy features, which were visible in Fig.~\ref{fig:effects}(a), are now highly entangled,
and the high-energy feature becomes very weak. The 3-spin interactions were crucial for explaining the magnetization
data, but as regards INS they actually have a counter productive effect. If it were not for the magnetization, one
would rather abandon them in an interpretation of the INS data. We chose $J_3 = -0.53$~K. Next the ZFS term was
included; the simulations for varying $D$ values are shown in Fig.~\ref{fig:effects}(c). As expected, the ZFS term has
a significant effect, and for $D$ values in between $-0.5 J$ and $-0.1 J$ both peaks I and II are well reproduced.
However, as discussed before, even for large $D$ values the mixing of spin levels is not strong enough to produce a
significant INS intensity at the higher energies, i.e., peak III is not reproduced. Furthermore, at 2.4~K there is
significant intensity in the energy region 0.8 to 1.2~meV, in contrast to experiment. We chose $D = -0.4 J = 3.5$~K.
Finally, the effect of the DM interactions on the INS spectrum at 9.3~K is shown in Fig.~\ref{fig:effects}(d).  The DM
interactions could in principle also be responsible for a mixing of spin levels and hence a violation of the INS
selection rule, but apparently they have no essential effect. We therefore conclude that DM interactions are not
relevant for explaining the key unexplained aspects of the magnetism in Ni$_4$Mo$_{12}$.

\begin{figure}
\begin{center}
\includegraphics{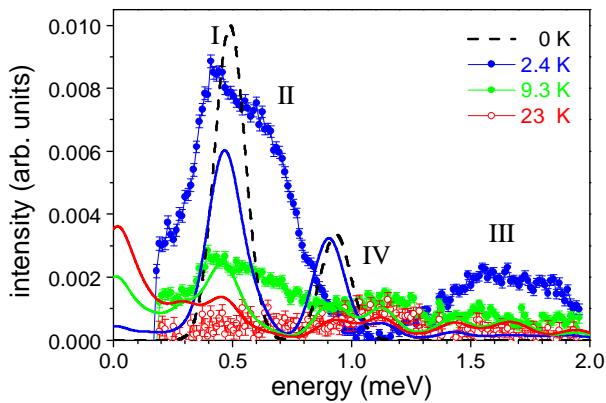}
\end{center}
\caption{
Experimental (dots) and simulated (lines) INS spectra for our best-fit model,
$\widehat{\mathcal H} = \widehat{\mathcal H}_{ex} + \widehat{\mathcal H}_{2} + \widehat{\mathcal H}_{3} +
\widehat{\mathcal H}_{ZFS}$ with $J = -8.82 \, \text{K}$, $J_2 = -1.08 \, \text{K}$, $J_3 = -0.56 \, \text{K}$, and $D = 3.5 \, \text{K}$.
}
\label{fig:bestd}
\end{figure}

The best simulation obtained in this approach is compared to the
experiment in Fig.~\ref{fig:bestd}. Peaks I and II are
reproduced reasonably well. Also the temperature dependence of
peak IV is by and large correctly obtained (increasing intensity
with increasing temperature). However, as for all models
discussed in this work, the intensity and temperature dependence
of peak III is not reproduced.

To conclude this section, all our efforts to reproduce the
experimental INS data with a spin Hamiltonian were unsuccessful
in the sense that key aspects of the data, such as the magnitude
of the intensity of peak III and its temperature dependence,
could not be satisfactorily reproduced. The analysis, however,
provided some insight into which terms should be most
important. For example, deviations from the tetrahedral symmetry
and DM interactions appear to be irrelevant.  Considering the
extensive yet unsuccessful spin-Hamiltonian-based efforts
presented here as well as in Ref.~\cite{Ni4Schnack}, one may
wonder about the apparent inability of the spin Hamiltonian
approach to account for the magnetism in Ni$_4$Mo$_{12}$.

\section{Hubbard Model Description}\label{sec:hubbard}

The difficulties in interpreting the experimental data by a spin
Hamiltonian suggest to reanalyze its basis. A possible approach
is to start with the more fundamental Hubbard model at
half-filling, from which spin exchange interactions are obtained
with a standard argument in a large $U$ expansion. In fact, a
Hubbard model was recently proposed for Ni$_4$Mo$_{12}$
\cite{Kost}. However, as we will show in this section, this
model is not adequate for describing the low-temperature
thermodynamic and spectroscopic properties of the molecule
either.

\begin{figure}
\begin{center}
\includegraphics[scale=0.75]{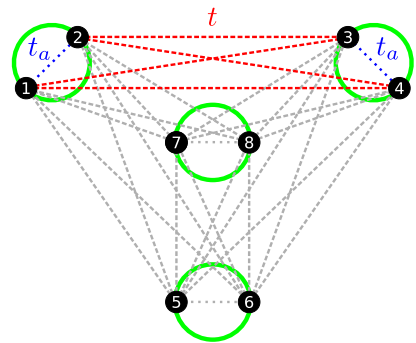}
\end{center}
\caption{
Schematic representation of the lattice and the hopping paths for the Hubbard model $\widehat{\mathcal{H}}_H$ (cf. Fig.~2 of Ref.~\cite{Kost}).
The small, numbered spheres correspond to Hubbard lattice sites, which are assigned to the four Ni ions as indicated by the large, green circles.
Possible jumps of electrons are visualized by lines: \emph{inter}-ion hoppings (dashed/red) occur with equal strength $t$ and
 \emph{intra}-ion hoppings (dotted/blue) with strength $t_a$, but the latter are excluded and set to zero.
}
\label{fig:hubbard_schema}
\end{figure}

As presented in Ref. \cite{Kost}, the motivation for applying a Hubbard model in the first place comes from the
octahedral surroundings of the Ni ions which cause a splitting of their $3d$ levels into three lower-lying $t_{2g}$ and
two higher-lying $e_g$ orbitals. According to Hund's rules, the $t_{2g}$ orbitals are fully occupied, whereas the $e_g$
orbitals are only singly occupied. The proposed Hubbard model is then formulated using a standard (one-band)
Hamiltonian \cite{Hubbard},

\begin{equation}
	\widehat{\mathcal{H}}_H = \sum_{\alpha\beta, \sigma}^{L}{t_{\alpha\beta} \hat{c}^{\dagger}_{\alpha,\sigma} \hat{c}_{\beta,\sigma}} + U \sum_{\alpha=1}^{L}{\hat{n}_{\alpha, \uparrow} \hat{n}_{\alpha, \downarrow}} + g \mu_{B} \hat{\vec{S}} \cdot \vec{B} \; ,
	\label{eq:hubbard_model}
\end{equation}
with parameters $N_e = L = 8$ (half-filling), where $L$ is the number of lattice sites and $N_e$ the number of electrons.
The $\hat{c}^{\dagger}_{\alpha, \sigma}$ ($\hat{c}_{\alpha, \sigma}$) are the fermionic creation (destruction)
operators for an electron at lattice site $\alpha$ with spin projection $\sigma = \, \uparrow, \downarrow$, and
$\hat{n}_{\alpha, \sigma} = \hat{c}^{\dagger}_{\alpha, \sigma} \hat{c}_{\alpha, \sigma}$. The local spin density at
site $\alpha$ can be written as $\hat{\vec{s}}_\alpha = \sum_{\sigma\tau}{\hat{c}^{\dagger}_{\alpha, \sigma}
\pmb{\sigma}_{\sigma \tau} \hat{c}_{\alpha, \tau}}$, with the vector of Pauli matrices $\pmb{\sigma}$. The total spin
operator then reads $\hat{\vec{S}}=\sum_{\alpha}{\hat{\vec{s}}_\alpha}$. The possible jumps of electrons between
lattice sites are characterized by the hopping parameters $t_{\alpha\beta}$, as illustrated in
Fig.~\ref{fig:hubbard_schema}. In the considered model all hopping matrix elements corresponding to \emph{inter}-ion
jumps are equal to $t$, and those corresponding to \emph{intra}-ion jumps are equal to $t_a$. The parameter $t_a$ is
set to zero, as in Ref. \cite{Kost}.

\begin{figure}
\hspace*{1pt} \includegraphics[scale=0.3, angle=-90]{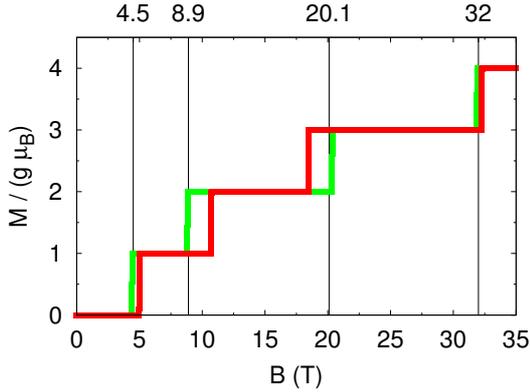}
\caption{
Magnetization of the Hubbard model $\widehat{\mathcal{H}}_H$ as a function of the applied magnetic field $B$ at $T = 0.01 \, \text{K}$
for the parameters $t=3.91$~meV, $U=64.6$~meV, and $g=2.25$ (red/dark). The steps occur at magnetic fields of 5.0, 10.7, 18.5, and 32.3 T.
For comparison, also the fit result of Ref.~\cite{Kost} is plotted (green/light), which was obtained with the Hamiltonian
$\widehat{\mathcal{H}}_{ex} + \widehat{\mathcal{H}}_2 + \widehat{\mathcal{H}}_3 + \widehat{\mathcal{H}}_B$ (see section \ref{sec:kostyuchenko}).
The positions of the experimentally determined crossing fields are indicated by vertical lines.
}
\label{fig:hubbard_fit}
\end{figure}

In this section we will focus on the steps in the magnetization curve as predicted by the Ha\-miltonian
$\widehat{\mathcal{H}}_H$.  At this point we should stress that in Ref.~\cite{Kost} not the Hubbard model itself was
analyzed, but the spin model $\widehat{\mathcal{H}}_{ex} + \widehat{\mathcal{H}}_2 + \widehat{\mathcal{H}}_3 +
\widehat{\mathcal{H}}_B$ [Eqs.~(\ref{eq:heisenberg})-(\ref{eq:3spin}), and (\ref{eq:zeeman})], which is supposed to
approximate the full Hamiltonian $\widehat{\mathcal{H}}_H$. In contrast, here we directly study the Hubbard model by
using numerical exact diagonalization to calculate the eigenvalues of $\widehat{\mathcal{H}}_H$. The parameters $t$ and
$U$ were then fitted to the experimentally determined crossing fields by employing a numerical minimization routine without
imposing any constraints on the parameter values.
Temperature was set to $T=0$, and the $g$ factor to $g=2.25$, as in Ref. \cite{Kost}. Despite testing a lot of initial
values and initial search directions for $t$ and $U$, we only found a single, unsatisfactory fit which obeys $U > 0$,
namely $t = 3.91$~meV and $U = 64.6$~meV. The predicted low-temperature magnetization curve is shown in
Fig.~\ref{fig:hubbard_fit}. The $\chi^2$ value of the fit to the crossing fields is about $6.25$~T$^2$, which is
approximately two orders of magnitude worse than the result obtained with the spin model $\widehat{\mathcal{H}}_{ex} +
\widehat{\mathcal{H}}_2 + \widehat{\mathcal{H}}_3 + \widehat{\mathcal{H}}_B$ \cite{Kost}. Considering the stability of
our fit result regardless of the starting conditions, and taking into account that the parameter space of the Hubbard
model $\widehat{\mathcal{H}}_H$ is only two-dimensional, we have to conclude that $\widehat{\mathcal{H}}_H$ cannot
explain the low-temperature magnetization in Ni$_4$Mo$_{12}$. Accordingly, it does not represent an adequate model for
Ni$_4$Mo$_{12}$, and thus does not capture the physical origin of the non-Heisenberg interactions in the molecule. This
clearly contradicts one of the main findings of Ref. \cite{Kost}. In the following we will hence substantiate our
conclusion and clarify the situation. As a byproduct, we will identify an error in the analytical derivation of the
spin model $\widehat{\mathcal{H}}_{ex} + \widehat{\mathcal{H}}_2 + \widehat{\mathcal{H}}_3 + \widehat{\mathcal{H}}_B$
in Ref. \cite{Kost}.

Before doing so, we should first comment on the proposed Hubbard model $\widehat{\mathcal{H}}_H$ and point out some of
its shortcomings and possible modifications. As far as we understand it, the \emph{intra}-ion hopping terms were
neglected in Ref. \cite{Kost} on the grounds that the Pauli principle prohibits a hopping between lattice sites which
are occupied by electrons with equal spin projection. However, this argument is valid only in a classical picture of a
spin-1 state. Quantum mechanically there is no \emph{a priori} reason for neglecting the \emph{intra}-ion hopping,
because a level with $s=1$ can also have a magnetic quantum number of $m=0$. Neglecting the \emph{intra}-ion hopping is
hence an inadequate method to incorporate the known ground-state spin configuration of the Ni ions into the Hubbard
model (see below). We have checked for a number of parameter sets that a non-zero value of $t_a$ can
influence the low-temperature magnetization curve by shifting the position of the ground state energies in
subspaces with fixed magnetic quantum number. In our fits to the magnetization data we therefore also added $t_a$ as a third independent fit
parameter. However, this did not lead to any improvement. Furthermore, according to the motivation for using a Hubbard
model as presented before, one should actually use a two-band Hubbard model because every Ni ion provides two magnetic
orbitals. In the usual derivation of the Hubbard model (see e.g. Ref. \cite{1DHubbard}) this leads to a variety of
additional interaction terms (\emph{intra}-ion Coulomb repulsion, \emph{intra}-ion exchange interaction, pair jumps,
and correlated hopping). Of these we considered only the terms which are usually studied in the literature, i.e.,
longer-range Coulomb repulsion and Heisenberg exchange (extended Hubbard model). The Heisenberg exchange term provides
a proper means to handle the spin state of the Ni ions, through the use of a ferromagnetic \emph{intra}-ion coupling
(Hund's rule coupling). Perhaps not surprisingly, the inclusion of further interaction terms allowed for a better fit
of the magnetization data and improved the $\chi^2$ value by about one order of magnitude. However, this is still one
order of magnitude worse than the fit result obtained in Ref.~\cite{Kost}. We thus do not believe that such
generalizations of the Hubbard model $\widehat{\mathcal{H}}_H$ lead in the right direction.

In order to resolve the apparent contradiction between our conclusion and that in Ref. \cite{Kost}, we reanalyzed the
strong-coupling limit $U \gg |t|$ of the Hamiltonian $\widehat{\mathcal{H}}_H$. In this limit the hopping term can be
treated as a perturbation, which at half-filling leads to an effective spin model $\widehat{\mathcal{H}}_{s}$ whose
energy eigenvalues are supposed to approximate the low-energy spectrum of the full Hubbard model \cite{1DHubbard}. It
should be noted that the effective Hamiltonian $\widehat{\mathcal{H}}_{s}$ is a spin-1/2 model, consisting of 8 spins
in the case of Ni$_4$Mo$_{12}$. We calculated $\widehat{\mathcal{H}}_{s}$ analytically up to order
$\mathcal{O}(t^4U^{-3})$ using two approaches. Starting from the results of the so called Canonical Transformation
\cite{tU,tU3}, Hubbard-X operators \cite{HubbardX} were employed to rewrite the effective Hamiltonian in the form of a
pure spin model, as in Ref.~\cite{tUHubbardX}. Alternatively, one can use the general results of Takahashi
\cite{Takahashi} and MacDonald \emph{et al.} \cite{tU2} to directly find the effective spin Hamiltonian, which
considerably speeds up the whole calculation. Both approaches lead to equivalent results. Denoting the spin Hamiltonian
up to order $\mathcal{O}(t^2 U^{-1})$ as $\widehat{\mathcal{H}}^{(2)}_s$, and the one up to order $\mathcal{O}(t^4
U^{-3})$ as $\widehat{\mathcal{H}}^{(4)}_s$, we obtained the following expressions:
\begin{subequations}\label{eq:effSpinModel}
\begin{eqnarray}
	\label{eq:2effective_spin_model}
  \widehat{\mathcal{H}}^{(2)}_s & =& -j^{(2)} \sum_{\alpha<\beta}^{8} {\hat{\vec{s}}_\alpha \cdot \hat{\vec{s}}_\beta}
   -  \frac{24 t^2}{U}+ g \mu_{B} \hat{\vec{S}} \cdot \vec{B} \; , \\
	\label{eq:4effective_spin_model}
  \widehat{\mathcal{H}}^{(4)}_s & = & -j^{(4)} \sum_{\alpha<\beta} ^{8}{\hat{\vec{s}}_\alpha \cdot \hat{\vec{s}}_\beta} - j_{ion} \sum_{i=1}^{4}{\hat{\vec{s}}_{\alpha_i} \cdot \hat{\vec{s}}_{\beta_i}} \\
	&& - j_4 \sum_{(\alpha\beta\gamma\delta)} \Big[ \left( \hat{\vec{s}}_\alpha \cdot \hat{\vec{s}}_\beta \right) \left( \hat{\vec{s}}_\gamma \cdot \hat{\vec{s}}_\delta \right)
        \nonumber \\
  && \qquad + \left( \hat{\vec{s}}_\alpha \cdot \hat{\vec{s}}_\delta \right) \left( \hat{\vec{s}}_\beta \cdot \hat{\vec{s}}_\gamma \right)
	  - \, \left( \hat{\vec{s}}_\alpha \cdot \hat{\vec{s}}_\gamma \right) \left( \hat{\vec{s}}_\beta \cdot \hat{\vec{s}}_\delta \right) \Big] \nonumber \\
	&& -  \left( \frac{24 t^2}{U} - \frac{78 t^4}{U^3} \right) + g \mu_{B} \hat{\vec{S}} \cdot \vec{B}  \nonumber \; .
\end{eqnarray}
\end{subequations}
The terms $-j^{(2)} \sum_{\alpha<\beta}^{8} {\hat{\vec{s}}_\alpha \cdot \hat{\vec{s}}_\beta}$ and $-j^{(4)}
\sum_{\alpha<\beta} ^{8}{\hat{\vec{s}}_\alpha \cdot \hat{\vec{s}}_\beta}$ describe the \emph{inter}-ion interactions,
where the sites $\alpha$ and $\beta$ belong to different ions; the term $- j_{ion}
\sum_{i=1}^{4}{\hat{\vec{s}}_{\alpha_i} \cdot \hat{\vec{s}}_{\beta_i}} $ describes the \emph{intra}-ion interactions,
where the sites $\alpha_i$ and $\beta_i$ belong to the same Ni ion $i$ [$(\alpha_i,\beta_i)$  = (1,2), (3,4), (5,6),
(7,8) for $i = 1, 2, 3, 4$]. $(\alpha\beta\gamma\delta)$ denotes the tetragon with vertices $\alpha$, $\beta$,
$\gamma$, and $\delta$, which have to be connected by hopping paths such that $\alpha$ is connected to $\beta$, $\beta$
to $\gamma$, $\gamma$ to $\delta$, and $\delta$ to $\alpha$ (see also Fig.~\ref{fig:hubbard_schema}). The Zeeman term
$g \mu_{B} \hat{\vec{S}} \cdot \vec{B}$ is unaffected in the perturbation theory as the total spin $\hat{\vec{S}}$
commutes with the hopping term. The coupling parameters were obtained as
\begin{subequations}\label{eq:effSpinParas}
\begin{eqnarray}
	\label{eq:2coupling_paras}
	j^{(2)} & = & -\frac{4 t^2}{U},\\
	\label{eq:4coupling_paras}
	j^{(4)} & = & - \frac{4 t^2}{U} + \frac{92 t^4}{U^3}, \:\:  j_4 = -\frac{80 t^4}{U^3}, \:\: j_{ion} = \frac{36 t^4}{U^3}.
\end{eqnarray}
\end{subequations}
Corrections to the spin Hamiltonian $\widehat{\mathcal{H}}^{(4)}_s$ and the given parameters are of order
$\mathcal{O}(t^6 U^{-5})$ \cite{Takahashi}. Even though \emph{intra}-ion hoppings are not included in the model, a
non-zero and ferromagnetic coupling between the spins belonging to the same Ni ion is obtained, $j_{ion}>0$. However,
it is of order $\mathcal{O}(t^4 U^{-3})$ and thus usually weak. We mention that the energy of the ferromagnetic ground
state with $|M| = |M_{max}|$ is zero for both $\widehat{\mathcal{H}}_H$ and its effective spin Hamiltonians [the
corresponding constants were explicitly retained in Eqs.~(\ref{eq:effSpinModel})].

\begin{figure}
\hspace*{0.5pt} \includegraphics[scale=0.325, angle=0]{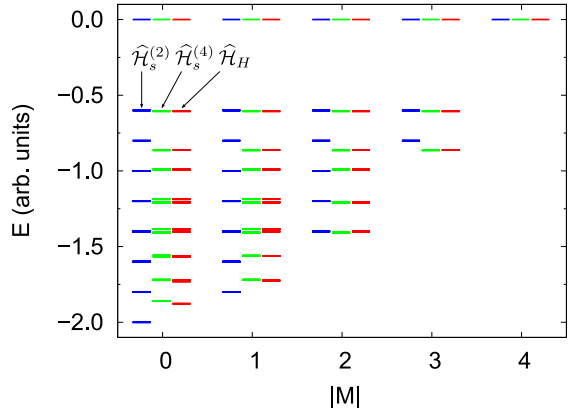}
\caption{
	Numerical comparison of the low-energy spectrum of the Hubbard model $\widehat{\mathcal{H}}_H$ (red/dark, right) and the complete spectrum of the spin model $\widehat{\mathcal{H}}_s^{(2)}$ (blue/dark, left) and $\widehat{\mathcal{H}}_s^{(4)}$ (green/light, middle) for the parameters $t=1$ and $U=20$. The energy eigenvalues are plotted vs. the absolute value of the total magnetic quantum number $|M|$. The product state with $|M|=4$ is an eigenstate of all three models with energy $E=0$.}
\label{fig:perturbation_spectra}
\end{figure}

In order to evaluate the accuracy of our analytical results for $\widehat{\mathcal{H}}_s^{(2)}$ and
$\widehat{\mathcal{H}}_s^{(4)}$, we calculated the energy eigenvalues of the different models for a number of parameter
sets $t$ and $U$ via numerical exact diagonalization and compared the obtained energy spectra. For the ratio $U/t=20$
the energy spectra are presented in Fig.~\ref{fig:perturbation_spectra} ($t=1$). Apparently, the spectrum of
$\widehat{\mathcal{H}}^{(2)}_s$ clearly deviates from the exact spectrum, in particular at low energies, whereas the
fourth-order Hamiltonian $\widehat{\mathcal{H}}^{(4)}_s$ gives a much better approximation. We found that for $U/t=30$ the low-energy spectrum of the Hubbard model $\widehat{\mathcal{H}}_H$ is already very well approximated by the
eigenvalues of $\widehat{\mathcal{H}}^{(4)}_s$. However, we also observed that for a ratio $U/t=10$, which was inferred
and identified with the strong-coupling limit $U \gg \, \rvert t \rvert$ in Ref. \cite{Kost}, the exact and
fourth-order spectrum have little in common. A link between exact and approximative levels may be established starting
with $U/t \approx 15$, but at this ratio clear discrepancies are still present, especially with respect to the ground
state energy. In the following we will only consider $\widehat{\mathcal{H}}^{(4)}_s$.

Having established an effective spin Hamiltonian for the
electron spins $\hat{\vec{s}}_\alpha$, we now discuss how to
derive an effective spin model in the space of the Ni spins
$\hat{\vec{S}}_i$. The total spin of the Ni ion $i$ is
$\hat{\vec{S}}_{i} = \hat{\vec{s}}_{\alpha_i} +
\hat{\vec{s}}_{\beta_i}$. Since the operators
$\hat{\vec{S}}_{i}$ commute with
$\widehat{\mathcal{H}}_s^{(4)}$, the quantum numbers $S_{i}$ are
good quantum numbers. The part of the energy spectrum with $S_i
= 1$ for all $i$ can thus be extracted by \emph{adding} a
ferromagnetic coupling between the electron spins
$\hat{\vec{s}}_{\alpha_i}$ and $\hat{\vec{s}}_{\beta_i}$
residing on the same Ni ion to $\widehat{\mathcal{H}}^{(4)}_s$
(Hund's rule coupling). Technically, this can be done by setting
$j_{ion} \rightarrow \infty$ (energy levels with $S_i
= 0$ for at least one $i$ are lifted to arbitrarily high
energies). However, it is important to note that this approach
is only possible within the effective spin model. In the case of
the Hubbard model $\widehat{\mathcal{H}}_{H}$ the $S_i$ are not
good quantum numbers since any operator which is specific to
particular Hubbard lattice sites, like $\hat{\vec{S}}_{i} =
\hat{\vec{s}}_{\alpha_i} + \hat{\vec{s}}_{\beta_i}$, cannot be a
conserved quantity in an itinerant model. Adding a Hund's rule
coupling directly to $\widehat{\mathcal{H}}_H$ would thus make
the configurations with $S_i=1$ energetically favorable, but it
would also change the structure of the whole energy
spectrum. This means that it is incorrect to state that the spin
model which is obtained by adding such a term to
$\widehat{\mathcal{H}}_s^{(4)}$ follows from the original
Hubbard model.

To carry out the limit $j_{ion} \rightarrow \infty$,
the operators $\hat{\vec{S}}_i$ can be interpreted as pure
spin-1 operators in a low-energy theory. The resulting spin
model then becomes (using notation introduced in section~3)
\begin{subequations}\label{eq:spin_model_transformed}
\begin{eqnarray}
	\label{eq:spin_model_transformed_a}
	\widehat{\mathcal{H}} &=& \widehat{\mathcal{H}}_{ex} + \widehat{\mathcal{H}}_2 + \widehat{\mathcal{H}}_3 + \widehat{\mathcal{H}}_B \\
   && -  J_4\sum_{(ijkl) \in \Gamma}{(\hat{\vec{S}}_i \cdot \hat{\vec{S}}_j) (\hat{\vec{S}}_k \cdot \hat{\vec{S}}_l)} -  \left( \frac{24 t^2}{U} + \frac{288t^4}{U^3} \right) \nonumber
\end{eqnarray}
with the coupling parameters
\begin{equation}
	\label{eq:paras_transformed}
	J = -\frac{4t^2}{U} + \frac{192t^4}{U^3}, \:\: J_2 = J_3 = \frac{J_4}{2} = -\frac{40 t^4}{U^3} \; ,
\end{equation}
\end{subequations}

and $\Gamma = \{(1,2,3,4),(1,3,2,4), (1,4,2,3)\}$. We mention that introducing an \emph{intra}-ion hopping of strength
$t_a$ = $t$ (instead of $t_a$ = 0) in $\widehat{\mathcal{H}}_{H}$ does not change the Hamiltonian
Eq.~(\ref{eq:spin_model_transformed}) (but certainly changes $\widehat{\mathcal{H}}_s^{(2)}$ and
$\widehat{\mathcal{H}}_s^{(4)}$), which is due to the high symmetry of the present case. These results finally clarify
our different findings for the magnetization steps when calculated with the Hubbard model $\widehat{\mathcal{H}}_H$ and
the spin model $\widehat{\mathcal{H}}_{ex} + \widehat{\mathcal{H}}_2 + \widehat{\mathcal{H}}_3 +
\widehat{\mathcal{H}}_B$, and should be compared to those in Ref. \cite{Kost}. One notable difference is that the
Hamiltonian Eq.~(\ref{eq:spin_model_transformed}) comprises 4-spin interactions. Starting from
Eq.~(\ref{eq:4effective_spin_model}), it can easily be verified that such terms have to appear in the projection of
$\widehat{\mathcal{H}}^{(4)}_s$ onto the subspace with $S_i = 1$ for all $i$. Furthermore, we find that the parameters
$J_2$ and $J_3$ are of equal strength, whereas in Ref. \cite{Kost} they differ by a factor of 2. Because the analysis
of the experimental magnetization data with the spin model $\widehat{\mathcal{H}}_{ex} + \widehat{\mathcal{H}}_2 +
\widehat{\mathcal{H}}_3 + \widehat{\mathcal{H}}_B$ yielded such a factor of 2 and the inferred ratio $U/t=10$, it was
concluded that the Hubbard model provides a valid description of the magnetism in Ni$_4$Mo$_{12}$ \cite{footnote1}.
However, our results demonstrate that such a conclusion is not supported.

\section{Conclusion}

In conclusion, we presented inelastic neutron scattering data
for the Ni$_4$Mo$_{12}$ molecule. An extensive analysis of these
data in terms of a phenomenological spin Hamiltonian did not
lead to a satisfactory description of its magnetism. A similar
observation was made in previous works, mainly based on
magnetization data \cite{Ni4Schnack}. Confirming this
observation by complementary spectroscopic data, in our opinion,
significantly furthers the idea that for Ni$_4$Mo$_{12}$ the
spin Hamiltonian approach is indeed inadequate. As an
alternative and more fundamental model the Hubbard model comes
to mind, which we hence analyzed in great detail. We have shown
that the first-guess Hubbard model $\widehat{\mathcal{H}}_H$
cannot explain the magnetization data of Ni$_4$Mo$_{12}$ as it
predicts incorrect crossing fields.  Obvious extensions of the
model did not resolve the issue satisfactorily either. We then
studied the strong-coupling limit of $\widehat{\mathcal{H}}_H$
in order to better understand these results. According to our
calculations, the observation in Ref.~\cite{Kost} that the
magnetization curve can be nicely fitted with the spin model
$\widehat{\mathcal{H}}_{ex} + \widehat{\mathcal{H}}_2 +
\widehat{\mathcal{H}}_3 + \widehat{\mathcal{H}}_B$ does not
imply that the Hubbard model is a suitable microscopic model for
Ni$_4$Mo$_{12}$. In this work we have in fact shown that the
opposite is true.

\section*{Acknowledgment}
We gratefully acknowledge financial support by the Deutsche Forschungsgemeinschaft and FOR 945.



\begin{thebibliography}{}

\bibitem{Mn12a}
  D. Gatteschi, R. Sessoli, J. Villain, \textit{Molecular Nanomagnets} (Oxford University Press, Oxford, 2006).

\bibitem{Mn12b}
  J. R. Friedman, M. P. Sarachik,  J. Tejada, R. Ziolo, Phys. Rev. Lett. \textbf{76}, 3830 (1996).

\bibitem{Mn12c}
  W. Wernsdorfer, R. Sessoli, Science \textbf{284}, 133 (1999).

\bibitem{BCC:IC99}
  J. J. Borras-Almenar, J. M. Clemente-Juan, E. Coronado,
  B. S. Tsukerblat, Inorg. Chem. \textbf{38}, 6081 (1999).

\bibitem{wheels1}
  K. L. Taft, C. D. Delfs, G. C. Papaefthymiou, S. Foner, D. Gatteschi, S. J. Lippard, J. Am. Chem. Soc. \textbf{116}, 823 (1994).

\bibitem{wheels2}
  D. Gatteschi, A. Caneschi, L. Pardi, R. Sessoli, Science \textbf{265}, 1054 (1994).

\bibitem{wheels2b}
 B. Pilawa, R. Desquiotz, M.T. Kelemen, M. Weickenmeier,
 A. Geisselman, J. Magn. Magn. Mater. \textbf{177}, 748 (1997).

\bibitem{wheels3}
  O. Waldmann, Coord. Chem. Rev. \textbf{249}, 2550 (2005).

\bibitem{wheels4}
  D. M. Low, G. Rajaraman, M. Helliwell, G. Timco, J. van Slageren, R. Sessoli, S. T. Ochsenbein, R. Bircher, C. Dobe, O. Waldmann, H. U. G\"{u}del, M. A. Adams, E. Ruiz, S. Alvarez, E. J. L. McInnes, Chem. Eur. J. \textbf{12}, 1385 (2006).

\bibitem{wheels5}
  R. E. P. Winpenny, Adv. Inorg. Chem. \textbf{52}, 1 (2003).

\bibitem{wheels6}
  A. L. Dearden, S. Parsons, R. E. P. Winpenny, Angew. Chem. Int. Ed. \textbf{40}, 151 (2001).

\bibitem{wheels6b}
 A. Cornia, M. Affronte, A. G. M. Jansen, G. L. Abbati, D. Gatteschi, Angew. Chem. Int. Ed. \textbf{38}, 2264 (1999).

\bibitem{wheels7}
  O. Waldmann, J. Sch\"{u}lein, R. Koch, P. M\"uller, I. Bernt, R. W. Saalfrank, H. P. Andres, H. U. G\"udel, P. Allenspach, Inorg. Chem. \textbf{38}, 5879 (1999).

\bibitem{wheels8}
  R. W. Saalfrank, I. Bernt, E. Uller, F. Hampel, Angew. Chem. Int. Ed. Engl. \textbf{36}, 2482 (1997).

\bibitem{wheels8b}
  J. van Slageren, R. Sessoli, D. Gatteschi, A. A. Smith, M. Helliwell, R. E. P. Winpenny, A. Cornia, A.-L. Barra, A. G. M. Jansen, E. Rentschler, G. A. Timco, Chem. Eur. J. \textbf{8}, 277 (2002).

\bibitem{wheels9}
  H. C. Yao, J. J. Wang, Y. S. Ma, O. Waldmann, W. X. Du, Y. Song, Y. Z. Li, L. M. Zheng, S. Decurtins, X. Q. Xin, Chem. Commun. \textbf{16}, 1745-1747 (2006).

\bibitem{wheels10}
  P. King, T. C. Stamatatos, K. A. Abboud, G. Christou, Angew. Chem. Int. Ed. Engl. \textbf{45}, 7379 (2006).

\bibitem{wheels11}
  K. B\"arwinkel, P. Hage, H.-J. Schmidt, J. Schnack, Phys. Rev. B \textbf{68}, 054422 (2003).

\bibitem{nrot1}
  O. Waldmann, T. Guidi, S. Carretta, C. Mondelli, A. L. Dearden, Phys. Rev. Lett. \textbf{91}, 237202 (2003).

\bibitem{nrot2}
  P. W. Anderson, \textit{Basic Notions of Condensed Matter Physics} (Benjamin/Cummings Publishing Co., Menlo Park, 1984).

\bibitem{nrot3}
  B. Bernu, C. Lhuillier, and L. Pierre, Phys. Rev. Lett. \textbf{69}, 2590 (1992).

\bibitem{waves1}
  O. Waldmann, Phys. Rev. B \textbf{65}, 024424 (2001).

\bibitem{nvt1}
  E. M. Chudnovsky, J. Tejada, \textit{Macroscopic Quantum Tunneling of the Magnetic Moment} (Cambridge University Press, Cambridge, 1998).

\bibitem{nvt2}
  B. Barbara, E. Chudnovsky, Phys. Lett. A \textbf{145}, 205 (1990).

\bibitem{nvt3}
  A. Chiolero, D. Loss, Phys. Rev. Lett. \textbf{80}, 169 (1998).

\bibitem{nvt4}
  O. Waldmann, T. C. Stamatatos, G. Christou, H. U. G\"{u}del, I. Sheikin, H. Mutka, Phys. Rev. Lett. \textbf{102}, 157202 (2009).

\bibitem{num1}
  D. Gatteschi, L. Pardi, Gazz. Chim. Ital. \textbf{123}, 231 (1993).

\bibitem{num2}
  O. Waldmann, Phys. Rev. B \textbf{61}, 6138 (2000).

\bibitem{fe30}
  A. M\"{u}ller, S. Sarkar, S. Q. N. Shah, H. B\"{o}gge, M. Schmidtmann, P. K\"{o}gerler, B. Hauptfleisch, A. Trautwein, V. Sch\"{u}nemann, Angew. Chem. Int. Ed. \textbf{38}, 3238 (1999).

\bibitem{KMS:CCR09}
  U. Kortz, A. M{\"u}ller, J. van Slageren, J. Schnack, N. S. Dalal, M. Dressel, Coord. Chem. Rev. \textbf{253}, 2315 (2009).

\bibitem{spinphase}
  C. Schr\"{o}der, H. Nojiri, J. Schnack, P. Hage, M. Luban, P. K\"{o}gerler, Phys. Rev. Lett. \textbf{94}, 017205 (2005).

\bibitem{RLM:PRB08}
  I. Rousochatzakis, A. M. L\"{a}uchli, F. Mila, Phys. Rev. B \textbf{77}, 094420 (2008)

\bibitem{SSR:JMMM05}
  R. Schmidt, J. Schnack, J. Richter, J. Magn. Magn. Mater. \textbf{295}, 164 (2005).

\bibitem{fe30bands}
  J. Schnack, M. Luban, R. Modler, Europhys. Lett. \textbf{56}, 863 (2001).

\bibitem{ExS:PRB03}
   M. Exler, J. Schnack, Phys. Rev. B \textbf{67}, 094440 (2003).


\bibitem{fe2a}
  R. Schenker, H. Weihe, H. U. G\"{u}del, Inorg. Chem. \textbf{40}, 4319 (2001).

\bibitem{fe2b}
  R. Schenker, M. N. Leuenberger, G. Chaboussant, H. U. G\"{u}del, and D. Loss, Chem. Phys. Lett. \textbf{358}, 413 (2002).

\bibitem{fe2c}
  R. Schenker, M. N. Leuenberger, G. Chaboussant, D. Loss, H. U. G\"{u}del, Phys. Rev. B \textbf{72}, 184403 (2005).

\bibitem{Cu4}
   O. Zaharko, J. Mesot, L. A. Salguero, R. Valent\'{\i}, M. Zbiri, M. Johnson, Y. Filinchuk, B. Klemke, K. Kiefer, M. Mys'kiv, Th. Str\"{a}ssle, H. Mutka, Phys. Rev. B \textbf{77}, 224408 (2008).

\bibitem{ni4smm1}
  A. Sieber, C. Boskovic, R. Bircher, O. Waldmann, S. T. Ochsenbein, G. Chaboussant, H. U. G\"{u}del, N. Kirchner, J. van Slageren, W. Wernsdorfer, A. Neels, H. Stoeckli-Evans, S. Janssen, F. Juranyi, H. Mutka, Inorg. Chem. \textbf{44}, 4315-4325 (2005).

\bibitem{ni4smm2}
  E. C. Yang, W. Wernsdorfer, L. N. Zakharov, Y. Karaki, A. Yamaguchi, R. M. Isidro, G. D. Lu, S. A. Wilson, A. L. Rheingold, H. Ishimoto, D. N. Hendrickson, Inorg. Chem. \textbf{45}, 529-546 (2006).

\bibitem{ni4smm3}
  E. del Barco, A. D. Kent, E. C. Yang, D. N. Hendrickson, Phys. Rev. Lett. \textbf{93}, 157202 (1997).

\bibitem{Ni4synthese}
  A. M\"{u}ller, C. Beugholt, P. K\"{o}gerler, H. B\"{o}gge, S. Budko, M. Luban, Inorg. Chem. \textbf{39}, 5176 (2000).

\bibitem{Ni4Schnack}
    J. Schnack, M. Br\"{u}ger, M. Luban, P. K\"{o}gerler, E. Morosan, R. Fuchs, R. Modler, H. Nojiri, R. C. Rai, J. Cao, J. L. Musfeldt, X. Wei, Phys. Rev. B \textbf{73}, 094401 (2006).

\bibitem{Kost}
  V. Kostyuchenko, Phys. Rev. B \textbf{76}, 212404 (2007).

\bibitem{Klemm}
  R. Klemm, D. Efremov, Phys. Rev. B \textbf{77}, 184410 (2008).

\bibitem{superINS1}
  G. Chaboussant, A. Sieber, S. Ochsenbein, H. U. G\"{u}del, M. Murrie, A. Honecker, N. Fukushima, B. Normand, Phys. Rev. B \textbf{70}, 104422 (2004).

\bibitem{superINS2}
  R. Bircher, G. Chaboussant, C. Dobe, H. U. G\"udel, S. T. Ochsenbein, A. Sieber, O. Waldmann,  Adv. Funct. Mater. \textbf{16}, 209 (2006).

\bibitem{superINS3}
  S. Carretta, P. Santini, G. Amoretti, T. Guidi, J. R. D. Copley, Y. Qiu, R. Caciuffo, G. Timco, R. E. P. Winpenny, Phys. Rev. Lett. \textbf{98}, 167401 (2007).

\bibitem{SFA:PRB07}
  M. B. Stone, F. Fernandez-Alonso, D. T. Adroja, N. S. Dalal,
  D. Villagr\'{a}n, F. A. Cotton, S. E. Nagler, Phys. Rev. B
  \textbf{75}, 214427 (2007).

\bibitem{WKD:JSSC03}
  M. H. Whangbo, H. J. Koo, D. Dai, J. Solid State
  Chem. \textbf{176}, 417 (2003).

\bibitem{MDM:CAEJ06}
  P. Mialane, C. Duboc, J. Marrot, E. Riviere, A. Dolbecq,
  F. Secheresse, Chem.-Eur. J. \textbf{12}, 1950 (2006).

\bibitem{HRZ:PRL05}
  M. Hagiwara, L. P. Regnault, A. Zheludev, A. Stunault,
  N. Metoki, T. Suzuki, S. Suga, K. Kakurai, Y. Koike,
  P. Vorderwisch, J. H. Chung, Phys. Rev. Lett.  \textbf{94},
  177202 (2005).




\bibitem{Hubbard}
  J. Hubbard, Proc. R. Soc. Ser. A \textbf{276}, 238 (1963).

\bibitem{tU}
  A. H. MacDonald, S. M. Girvin, D. Yoshioka, Phys. Rev. B \textbf{37}, 9753 (1988).

\bibitem{FNL:JACS96}
  J. A. Farrar, F. Neese, P. Lappalainen, P. M. H. Kroneck,
  M. Saraste, W. G. Zumft, A. J. Thomson,
  J. Am. Chem. Soc. \textbf{118}, 11501 (1996).



\bibitem{PhdBrueger}
  M. Br\"{u}ger, Ph.D. thesis, Osnabr\"{u}ck University (2008).

\bibitem{DM1}
  I. Dzyaloshinsky, J. Phys. Chem. Solids \textbf{4}, 241 (1958).

\bibitem{DM2}
  T. Moriya, Phys. Rev. \textbf{120}, 91 (1960).

\bibitem{Ni4Kirchner}
  N. Kirchner, J. van Slageren, B. Tsukerblat, O. Waldmann, M. Dressel, Phys. Rev. B \textbf{78}, 094426 (2008).

\bibitem{eulerbook}
  A. Messiah, \textit{Quantenmechanik} (Walter de Gruyter, Berlin, New York, 1985)

\bibitem{INS1}
O. Waldmann, Phys. Rev. B \textbf{68}, 174406 (2003).

\bibitem{INS2}
O. Waldmann, Phys. Rev. B \textbf{72}, 094422 (2005).

\bibitem{Boca}
R. Boca, Coord. Chem. Rev. \textbf{248}, 757 (2004).

\bibitem{Liviotti}
 E. Liviotti, S. Carretta, G. Amoretti, J. Chem. Phys. \textbf{117}, 3361 (2002).

\bibitem{Datta}
S. Datta, O. Waldmann, A. D. Kent, V. A. Milway, L. K. Thompson, S. Hill, Phys. Rev. B \textbf{76}, 052407 (2007).

\bibitem{1DHubbard}
F. H. Essler, H. Frahm, F. G\"{o}hmann, A. Kl\"{u}mper, V. E. Korepin, \textit{The One-Dimensional Hubbard Model} (Cambridge University Press, 2005).

\bibitem{tU3}
A. L. Chernyshev, D. Galanakis, P. Phillips, A. V. Rozhkov, A.-M. S. Tremblay, Phys. Rev. B \textbf{70}, 235111 (2004).

\bibitem{HubbardX}
J. Hubbard, Proc. Roy. Soc. London \textbf{285}, 542 (1965).

\bibitem{tUHubbardX}
P. Kakashvili, G. I. Japaridze, J. Phys.: Cond. Mat. \textbf{16}, 5815 (2004).

\bibitem{Takahashi}
M. Takahashi, J. Phys. C \textbf{10}, 1289 (1977).

\bibitem{tU2}
  A. H. MacDonald, S. M. Girvin, D. Yoshioka, Phys. Rev. B \textbf{43}, 6209 (1991).

\bibitem{footnote1}
  At this point it is important to note that the Hamiltonian
  (\ref{eq:spin_model_transformed}) and the Hubbard model
  $\widehat{\mathcal{H}}_H$ share the ground state energies in
  the subspaces with given $M$ in the interesting parameter
  regime, i.e., in the regime where
  $\widehat{\mathcal{H}}_s^{(4)}$ can be an approximation for
  $\widehat{\mathcal{H}}_H$. For this reason they should
  predict the same crossing fields at $T=0$ in this
  regime. So, while the projection onto the space with $S_i =
  1$ for all $i$ in general breaks the connection between the
  effective spin model and the full Hubbard model, it does
  actually not matter in case of the magnetization at $T=0$.

\end{thebibliography}
\end{document}